%% file: main.tex
\documentclass[sigconf]{acmart}

\AtBeginDocument{%
  \providecommand\BibTeX{{%
    \normalfont B\kern-0.5em{\scshape i\kern-0.25em b}\kern-0.8em\TeX}}}

\setcopyright{acmlicensed}
\copyrightyear{2018}
\acmYear{2018}
\acmDOI{XXXXXXX.XXXXXXX}

\acmConference[Conference acronym 'XX]{Make sure to enter the correct
  conference title from your rights confirmation emai}{June 03--05,
  2018}{Woodstock, NY}
\acmISBN{978-1-4503-XXXX-X/18/06}

\input{comments} % added by JooYoung to implement comment features.

\copyrightyear{2025}
\acmYear{2025}
\setcopyright{rightsretained}
\acmConference[ASSETS '25]{The 27th International ACM SIGACCESS Conference on Computers and Accessibility}{October 26--29, 2025}{Denver, CO, USA} \acmBooktitle{The 27th International ACM SIGACCESS Conference on Computers and Accessibility (ASSETS '25), October 26--29, 2025, Denver, CO, USA} \acmDOI{10.1145/3663547.3759765} \acmISBN{979-8-4007-0676-9/2025/10}

\begin{document}

\title[Supporting Multimodal Interaction with 3D Surface and Point Data Visualizations]{Explore, Listen, Inspect: Supporting Multimodal Interaction with 3D Surface and Point Data Visualizations}

\author{Sanchita S. Kamath}
\affiliation{%
 \department{School of Information Sciences}
  \institution{University of Illinois Urbana-Champaign}
  \orcid{0000-0001-6469-0360}
  \city{Champaign}
  \state{Illinois}
  \country{USA}
  \postcode{61820}
  }
\email{ssk11@illinois.edu}

\author{Aziz Zeidieh}
\orcid{0009-0000-9334-8660}
\affiliation{%
 \department{Informatics}
  \institution{University of Illinois Urbana-Champaign}
  \city{Champaign}
  \state{Illinois}
  \country{USA}
  \postcode{61820}
  }
\email{azeidi2@illinois.edu}

\author{JooYoung Seo}
\orcid{0000-0002-4064-6012}
\affiliation{%
 \department{School of Information Sciences}
  \institution{University of Illinois Urbana-Champaign}
  \city{Champaign}
  \state{Illinois}
  \country{USA}
  \postcode{61820}
  }
\email{jseo1005@illinois.edu}

\renewcommand{\shortauthors}{Kamath et al.}

\begin{abstract}
  \input{chapter/0_abstract}
\end{abstract}




\received{20 February 2007}
\received[revised]{12 March 2009}
\received[accepted]{5 June 2009}

\maketitle

\input{chapter/1_introduction}
\input{chapter/2_related_work}
\input{chapter/3_design_procedures}
\input{chapter/4_system_implementation}
\input{chapter/5_findings-discussion}
\input{chapter/6_conclusion}

% \begin{acks}
%   \input{chapter/7_acknowledgement}
% \end{acks}

\bibliographystyle{ACM-Reference-Format}
\bibliography{references/references}

\input{chapter/8_appendix}

\end{document}

%% file: comments.tex
\usepackage{todonotes}
\usepackage{soul}  % For highlighting text with \hl

%% file: chapter/0_abstract.tex
% 150 words. Must include the following components:
% 1. Issue, problem, gap, deficiency
% 2. Purpose (the participants who will be studied, and the site where the research will take place)
% 3. Data collection (the type of data, the participants, and where the data will be collected)
% 4. Results
% 5. Implications for target audience
Blind and low-vision (BLV) users remain largely excluded from three-dimensional (3D) surface and point data visualizations due to the reliance on visual interaction. Existing approaches inadequately support non-visual access, especially in browser-based environments. This study introduces \textit{\textbf{DIXTRAL}}, a hosted web-native system, co-designed with BLV researchers to address these gaps through multimodal interaction. Conducted with two blind and one sighted researcher, this study took place over sustained design sessions. Data were gathered through iterative testing of the prototype, collecting feedback on spatial navigation, sonification, and usability. Co-design observations demonstrate that synchronized auditory, visual, and textual feedback, combined with keyboard and gamepad navigation, enhances both structure discovery and orientation. \textbf{\textit{DIXTRAL}} aims to improve access to 3D continuous scalar fields for BLV users and inform best practices for creating inclusive 3D visualizations.

%% file: chapter/1_introduction.tex
\section{Introduction}
\label{sec:introduction}

% 1. State the research problem (fact-based, objective stats on the research topic).
Three-dimensional (3D) data visualization reveals spatial relationships, depth cues, and volumetric patterns \cite{papachristodoulou_sonification_2014} that two-dimensional (2D) charts cannot effectively display, making it invaluable in fields such as meteorology \cite{rautenhaus_three-dimensional_2015}, biomedical imaging \cite{zhou_review_2022}, VUV spectroscopy \cite{kaplitz_gas_2023}, geoscience \cite{qi_3d_2007}, and computational fluid dynamics \cite{harte_second_2024}.
% 2. Review prior studies that have addressed the problem (some citations).
Without 3D visualizations, analysts risk missing essential insights in spatially complex datasets \cite{bui_role_2021, eilola_3d_2023}. 
Specifically, surface plots are 3D visualizations of functions with two independent variables, where the X-Z plane forms the horizontal surface and the Y-axis represents the dependent variable's value. These plots are critical in aforementioned fields as they help reveal peaks, valleys, and gradients in spatial data \cite{papachristodoulou_sonification_2014}. 
% 3. Indicate deficiencies in the studies.
However, surface plots remain largely inaccessible to blind and low-vision (BLV) individuals, who often rely on screen readers and text-based tools that cannot convey 3D spatial features and have expressed a strong desire for richer exploratory interpretation tools \cite{keilers_data_2023}. Existing alternatives (e.g., alt-text or raw data tables) may increase cognitive load and slow comprehension \cite{keilers_data_2023}; and while non-visual access through haptic graphics \cite{paneels_review_2010} and 3D-printed physicalizations \cite{braier_haptic_2014} offer another route, they demand special hardware and are impractical to scale broadly \cite{vidal-verdu_graphical_2007}. Therefore, these efforts fall short in three areas: (1) surface plots remain largely inaccessible \cite{braier_haptic_2014}; (2) multimodal integration across vision, sound, and text is rarely coordinated \cite{sardana_multi-modal_2021, enge_open_2024}; and (3) browser-based implementations that support accessible, interactive 3D analysis are virtually nonexistent \cite{de_paiva_guimaraes_immersive_2018}.
% 4. Advance the significance of the study for particular audiences.
Addressing these limitations is imperative. Accessible surface visualization broadens STEM inclusion by enabling BLV users to independently explore spatial data \cite{jiang_designing_2024}.
% 5. State the purpose statement (one purpose).
The purpose of this study is to develop a web-native system for accessible surface data visualization, integrating real-time sonification, textual feedback, visual highlight, and structured spatial navigation. We focus on the web platform because it aligns with existing usage patterns of BLV users, who commonly rely on screen readers and are already familiar with navigating web-based content. Following are the research questions our study aims to answer:
% 6. State research questions (up to three).
\begin{enumerate}
    \item [\textbf{RQ1.}] How can surface-based 3D data be made accessible through multimodal interaction techniques?
    \item [\textbf{RQ2.}] How can interaction architecture support both whole-to-part and part-to-whole exploration of three-dimensional scalar fields for BLV users within a single interactive snapshot?
    \item [\textbf{RQ3.}] In what ways can an accessibility-first design support both BLV and sighted users through shared, non-conflicting multimodal features?
\end{enumerate}
% 7. State contributions of the study.
We present \textit{\textbf{DIXTRAL}} (Dynamic Interface eXploration of Three-dimensional Representations for Adaptive Learning) \footnote{https://sanchitakamath.com/dixtral-assets2025-demo/} as a proof-of-concept addressing our proposed RQs. \textbf{\textit{DIXTRAL}} is an accessible surface plot system that employs coordinated sonification alongside visual highlighting and textual description to support both part-to-whole and whole-to-part analysis. This approach builds on the idea that coordinated modalities enhance inclusive sensemaking \cite{albouys-perrois_towards_2018}. \textbf{\textit{DIXTRAL}} integrates surface data, multisensory encoding, and BLV accessibility into one coherent interface. The presented system contributes toward this gap by extending known 2D sonification techniques to browser-based 3D scalar field visualization, enabling equitable, multimodal data exploration. This work is inspired by~\citet{seo_maidr_2024}; we extend their approach to two-dimensional accessible data visualization by considering planar 2D slices of 3D data with tri-axis navigation and real-time auditory, textual, and visual alignment in a web-native interface.
% 8. Outline the structure of the paper.

%% file: chapter/2_related_work.tex
\section{Related Work}
\label{sec:related_work}
Prior work explored tactile representations of 2D charts \cite{paneels_review_2010, zhao_effectiveness_2020} and bar/pie charts via raised lines or 3D printing \cite{braier_haptic_2014}, while systems like HITPROTO \cite{paneels_prototyping_2013} and early force-feedback devices \cite{fritz_design_1999} enabled non-visual surface exploration. Sonification, in particular, within the 2D visualization space has been well-studied: \citet{walker_universal_2010} demonstrated pitch‑time mappings for line charts that allowed blind users to discern trends and slopes; \citet{ahmetovic_audiofunctionsweb_2019} provided real-time audio feedback for equation graphs; \citet{wang_seeing_2022} explored sonification of scatterplots through pitch, stereo pan, and onset variation to encode point distributions; and \citet{chundury_tactualplot_2024} applied sensory substitution to auditory-spatial navigation in 2D charts via touchscreens. 
However, the challenge of making 3D visualizations, commonly used to parallelly represent multiple continuous variables - such as precipitation, temperature and time - accessible remains underexplored.
Nevertheless, recent studies begin to address this: \citet{herman_touching_2025} found tactile models often more intuitive than digital or VR versions, and \citet{schwarz_interface_2022} proposed a surface-accessibility framework for simpler contexts. Though physicalization supports accessibility \cite{zhao_embodiment_2008}, most examples are static. Interactive or multimodal support for surface plot exploration remain rare. While embodied learning strategies show promise \cite{chen_be_2018, zhou_embodied_2024, trajkova_move_2020}, sonification, despite its demonstrated benefits for BLV scientists \cite{gomez_valencia_computer-vision_2014}, is rarely used in educational tools for 3D data. When aligned with navigable axes (or a three dimensional virtual grid), sound can effectively convey structural information \cite{lyon_auditory_1996, krygier_chapter_1994}, making it a powerful yet underutilized tool for representing 3D data in accessible educational contexts.
% Immersive systems
Further, immersive systems such as DXR \cite{sicat_dxr_2019}, VirtualDesk \cite{wagner_filho_virtualdesk_2018}, and CompositingVis \cite{zhu_compositingvis_2025} demonstrate how spatial reasoning enhances multivariate data exploration in VR. Others explore embodied navigation and perceptual-motor alignment \cite{yang_embodied_2021, lee_shared_2021, cordeil_imaxes_2017}. Yet in the browser, accessible tools for 3D scalar fields remain virtually nonexistent. Despite advances in WebGL \cite{pullan_visualization_nodate}, most web-native systems prioritize visual interaction, excluding users who rely on alternative modalities.

%% file: chapter/3_design_procedures.tex
\section{Design Procedures and Goals}
\label{sec:design_procedures_and_goals}

\subsection{Design Procedures}
\label{sec:design_procedures}
We used a mixed-ability co-design methodology grounded in the interdependence framework~\cite{bennett_interdependence_2018}, emphasizing shared authority and sustained collaboration across visual acuities. Our team comprised three researchers: two legally blind (R1 and R2) and one non-BLV (R3). R1, a screen reader user for over 24 years with extensive experience in auditory interaction, and R2, a user for 18 years, provided design insight, lived experience, and conducted early testing throughout the project. R3 led implementation and interface development, directly informed by R1 and R2’s feedback. Visual acuity profiles for all contributors are listed in Table~\ref{tab:vis_acuity}.

\begin{table}[H]
  \begin{center}
  \tiny
    \begin{tabular}{ p{1.5cm} p{6cm} }
      \hline
      \textbf{ResearcherID} & \textbf{Visual Acuity} \\
      \hline
      R1 & Blind with no perception of lights and shapes in the right eye and ability to perceive lights and shapes in the left. \\ 
      \hline
      R2 & Blind with no perception of lights and shapes in the left eye and bilateral retinal detachment with acuity of 20/2000 in the right eye. \\  
      \hline
      R3 & Not BLV, diagnosed with myopia, wears corrective lenses while testing. \\
      \hline
    \end{tabular}
    \caption{Visual Acuity of Researchers}
    \label{tab:vis_acuity}
  \end{center}
\end{table}

Development of \textbf{\textit{DIXTRAL}} began in May 2025, followed by a month of sustained co-design. R3 met weekly with R1 and collaborated daily with R2. Each session involved evaluating the prototype, proposing new features, and addressing interaction pain points, with R3 implementing refinements iteratively. Drawing on expertise in screen reader use, web development, audio production, and accessibility tooling, we reframed sighted affordances, such as highlight and grid overlays, into non-visual strategies like wireframe traversal, multimodal alignment, and axis-specific navigation.

\subsection{Design Goals}
\label{sec:design_goals}
% through design process, identify design goals.
% up to 5 design goals here... Each design goal is denoted with DG and use "should" verb. Each goal needs to be concisely supported by evidence/prior work.
Inspired by our co-design sessions, we defined the following system-level goals:
\begin{itemize}
\item \textbf{DG1: The system should convey 3D spatial structure through coordinated sonification and visual highlighting.}
R2 emphasized the difficulty of interpreting spatial depth through visuals alone. Our co-design sessions showed that combining sonification with visual highlighting improved orientation and spatial reasoning. This aligns with prior work demonstrating that sound can effectively convey spatial features via pitch and timbre \cite{bizley_interdependent_2009}.
\item \textbf{DG2: The system should enable spatial navigation across all axes via discrete input mappings.}
Our BLV co-designers expressed a preference for discrete navigation over continuous input, citing greater control and predictability. This supports previous findings that mouse-based 3D interaction excludes many non-visual users \cite{hand_survey_1997}.
\item \textbf{DG3: The system should expose surface data through adaptable verbal and textual descriptions.}
Research shows that context-aware textual feedback improves understanding of complex visuals \cite{meyer_accessibility_2023}. This aligned with the feedback from our co-designers highlighting the need for layered descriptions - quick summaries, with the option to drill down into more detail.
\item \textbf{DG4: The system should maintain modular architecture with strict separation of concerns.}
Our development process surfaced the need to maintain accessibility features independently of rendering logic to support experimentation and long-term sustainability. A modular design would be able to support long-term sustainability and easier integration of new accessibility techniques.
\item \textbf{DG5: The system should support heterogeneous datasets and user uploads.}
R1 requested the ability to explore their own data, from diverse domains. Supporting this flexibility aligns with inclusive design principles and would increase the tool’s real-world applicability.
\end{itemize}

%% file: chapter/4_system_implementation.tex
\section{System Implementation}
\label{sec:system_implementation}

\begin{figure}[ht]
\centering
\includegraphics[width=1\linewidth]{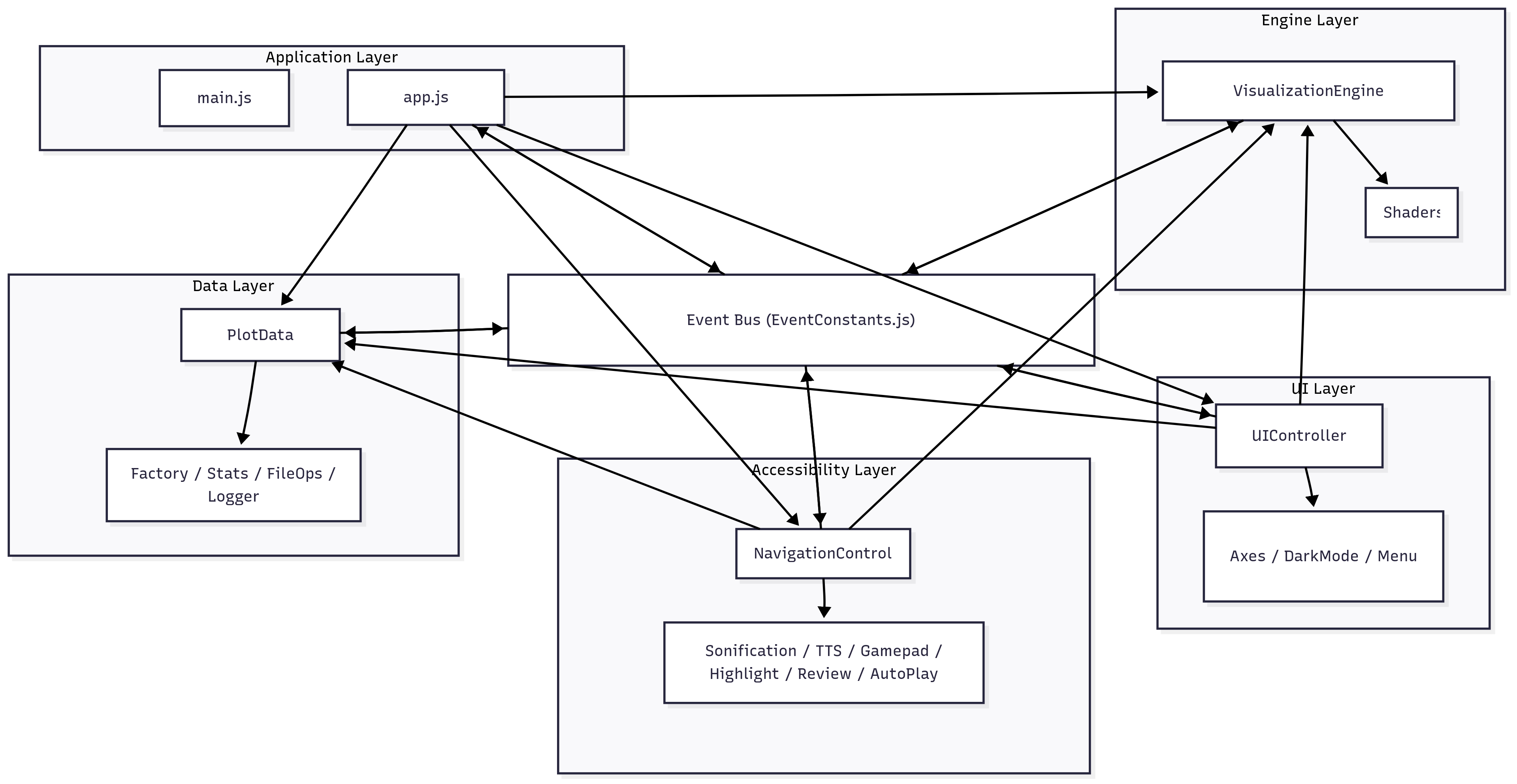}
\caption{System Diagram}
\Description{A top-down diagram showing the event-driven architecture of DIXTRAL, divided into five boxed layers: Application, UI, Accessibility, Engine, and Data. Each box contains key modules, and arrows show communication flow. The Application Layer includes main.js and app.js, which point to all other layers. At the center is a horizontal “Event Bus (EventConstants.js)” box that connects bidirectionally with app.js, UIController, NavigationController, VisualizationEngine, and PlotData. Each controller module in the UI and Accessibility layers connects to their internal components: UIController to “Axes / DarkMode / Menu”, and NavigationController to “Sonification / TTS / Gamepad / Highlight / Review / AutoPlay”. The Engine Layer has VisualizationEngine pointing to Shaders, and the Data Layer includes PlotData connected to “Factory / Stats / FileOps / Logger”.}
\label{fig:sys-dia}
\end{figure}

\begin{figure}[ht]
\centering
\includegraphics[width=1\linewidth]{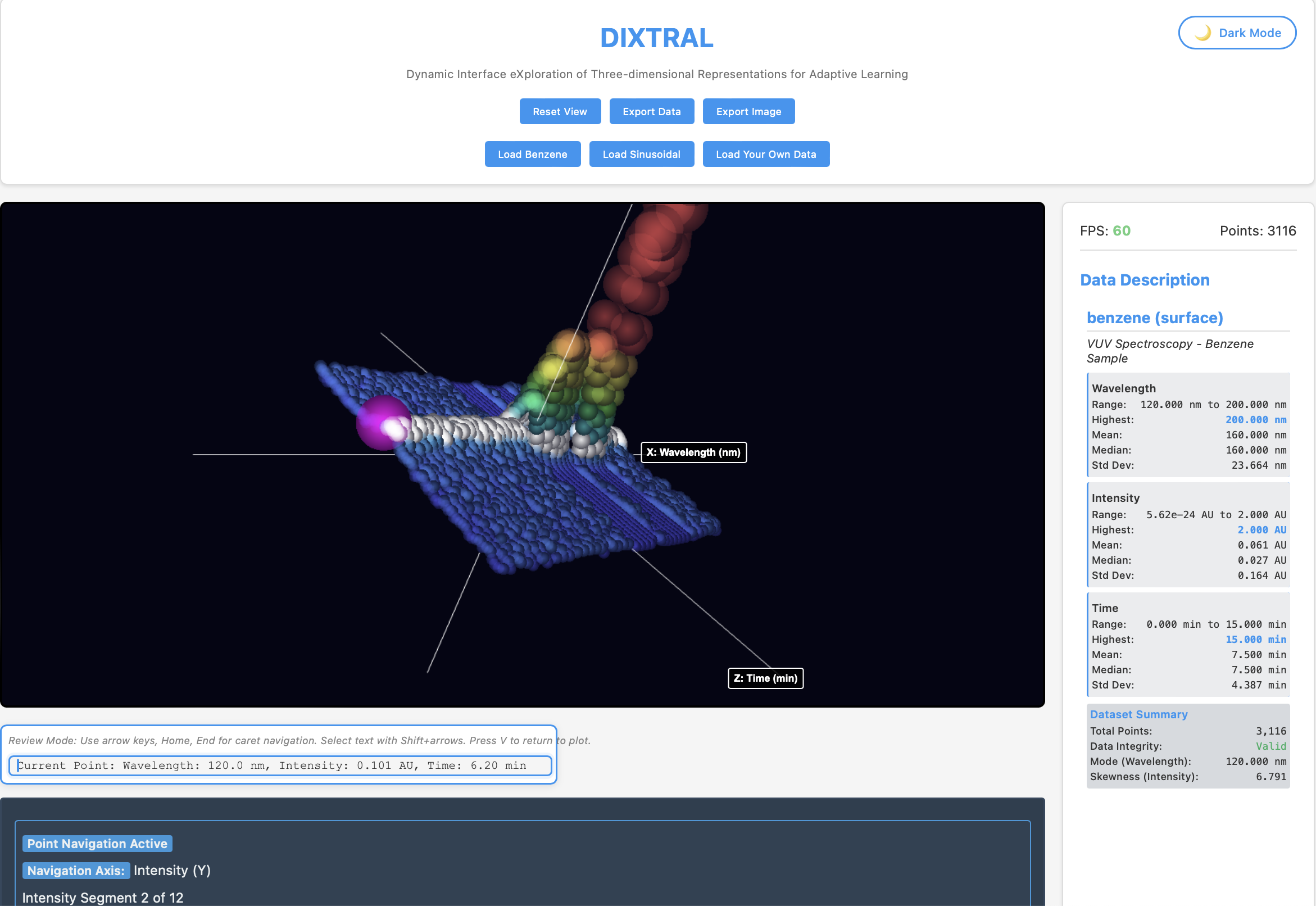}
\Description{A web interface titled ``DIXTRAL: Dynamic Interface eXploration of Three-dimensional Representations for Adaptive Learning''. At the top are buttons: Dark Mode, Reset View, Export Data, Export Image, Load Benzene, Load Sinusoidal, and Load Your Own Data. Below is a 3D plot showing a multicolored surface with a vertical cluster of red, orange, yellow, green and blue spheres, representing intensity through color. There is one segment along the Y axis that has been highlighted in white and one point within that has been highlighted in magenta. A label points to axes labeled ``X: Wavelength (nm)'', ``Y: Intensity (AU)'', and ``Z: Time (min)''. To the right, a sidebar titled ``Data Description'' shows metadata for the loaded ``benzene (surface)'' sample. It lists statistical summaries for three dimensions: Wavelength, Intensity, and Time — including range, mean, median, and standard deviation. The dataset contains 3,116 points, with a data integrity status marked ``Valid''. Below the plot, a review mode box displays the current data point: Wavelength = 120.0 nm, Intensity = 0.101 AU, Time = 6.20 min. Under that, a panel shows navigation mode status: ``Point Navigation Active'', axis set to ``Intensity (Y)'', and indicates ``Segment 2 of 12''.}
\caption{\textbf{\textit{DIXTRAL}} Interface}
\label{fig:dixtral-interface}
\end{figure}

\textbf{\textit{DIXTRAL}} is a browser-based system designed to facilitate structured, multimodal exploration of three-dimensional data. Built with accessibility as a first-class principle, it offers users interaction via synchronized visual, auditory \textit{(DG1)}, and textual modalities \textit{(DG3)}. DIXTRAL uses a modular, event-driven architecture (visualized in Figure \ref{fig:sys-dia}) with a custom WebGL engine that supports dynamic buffer generation, live shader management, and performance-optimized redraws. The engine manages the rendering of point clouds, wireframe grids, and surface meshes in real time \textit{(DG1, DG2)}. All interactions are synchronized with accessibility events via a decoupled architecture that ensures strict separation of concerns \textit{(DG4)}. The user interface is viewable in Figure~\ref{fig:dixtral-interface}.

% The \urlstyle and \url commands require \usepackage{url} or \usepackage{hyperref} in your document preamble.
% The \sloppy command is added to further prevent text from overflowing the column boundaries.

\sloppy 

\textbf{\textit{DIXTRAL}} adopts a strict five-layer modular architecture to enforce a clean separation of concerns, where each layer has a distinct purpose and communicates through designated coordinators. This design ensures maintainability, testability, and extensibility. The five layers are as follows:

\begin{itemize}
    \item \textbf{Application Layer:} This top-level layer is responsible for global coordination and the application's lifecycle management. Coordinated by \urlstyle{tt}\url{app.js} and bootstrapped by \urlstyle{tt}\url{main.js}, its primary responsibilities are to initialize all other layer coordinators in the correct dependency order to ensure system stability, handle global event coordination, and provide global error handling mechanisms \textit{(DG5)}.
    
    \item \textbf{UI Layer:} This layer manages all user interface components and visual controls. It is orchestrated by the \urlstyle{tt}\url{UIController.js}, which manages a suite of sub-components including the \urlstyle{tt}\url{AxesController} for rendering 3D axes, the \urlstyle{tt}\url{DarkModeController} for theme management, and the \urlstyle{tt}\url{MenuController} for the help system. Its responsibilities include handling all visual updates, coordinating the plot type selection interface, and managing user-facing controls like the TTS toggle.
    
    \item \textbf{Accessibility Layer:} This layer centralizes all non-visual access and assistive functionalities, making the platform universally accessible. It is coordinated by the \urlstyle{tt}\url{NavigationController.js}, which manages a comprehensive set of components like the \urlstyle{tt}\url{SonificationController}, \urlstyle{tt}\url{TextController}, \urlstyle{tt}\url{TTSController}, \urlstyle{tt}\url{GamepadController}, \urlstyle{tt}\url{HighlightController}, and \urlstyle{tt}\url{ReviewModeController}. This layer's key responsibilities include managing a tri-axis navigation system (with Y, Z, and X modes), providing multi-dimensional audio feedback through sonification, and handling both point and filled-rectangle highlighting. It also implements "Review Mode" for seamless focus switching for screen reader users and supports dual navigation strategies: a detailed "Part-to-Whole" exploration and a rapid "Whole-to-Part" overview via an intelligent autoplay system \textit{(DG2, DG3)}.
    
    \item \textbf{Engine Layer:} This layer contains the WebGL-based rendering pipeline, which is responsible for all visualization tasks. The \urlstyle{tt}\url{VisualizationEngine.js} acts as the coordinator, managing the entire rendering process from shader management to performance optimization. It provides a universal rendering solution that can support all plot types, ensuring that visual effects and rendering are handled efficiently and consistently \textit{(DG1, DG2)}.
    
    \item \textbf{Data Layer:} This layer manages all data-related operations, including ingestion, generation, and analysis. Its central hub is \urlstyle{tt}\url{PlotData.js}, which works with a \urlstyle{tt}\url{PlotDataFactory.js} to enable modular support for multiple plot types through a factory pattern. This layer's responsibilities include central data storage, validation, file import/export operations, and data format conversions. Furthermore, it integrates a comprehensive descriptive statistics engine that calculates measures such as mean, median, mode, standard deviation and variance for all data dimensions \textit{(DG5)}.
\end{itemize}

This architecture is governed by a critical rule: cross-layer communication is permitted only between designated coordinator files. All other components must communicate through their respective coordinators, primarily using a centralized, event-driven protocol or dependency injection \textit{(DG4)}. This design preserves architectural integrity and ensures that accessibility systems can operate independently of the rendering engine, facilitating future extensions without requiring foundational code rewrites.

\subsection{Rendering and Navigation}
\label{subsec:rendering-nav}
\textbf{\textit{DIXTRAL}} offers two primary display modes rendered via a shared WebGL pipeline: a Surface Mode, which constructs a topological mesh with a wireframe grid overlay, and a Point Mode, which displays individual data coordinates as a point cloud, ideal for sparser datasets \textit{(DG1, DG2)}. The user can toggle the visibility of the 3D axes and their labels with the `A` key. All keybindings are summarized in Table~\ref{tab:dixtral-keybindings}.

The system's tri-axis navigation defines a virtual grid segmented along the X, Y, or Z axes. A user selects the primary navigation axis using the `N` key and moves point-by-point with the arrow keys. For more rapid traversal, `Ctrl+Shift+arrow` combinations jump between segments, which are defined by grouping points with similar coordinate values along the selected axis \textit{(DG2)}. This architecture supports two complementary navigation strategies:
\begin{itemize}
    \item [1.] \textit{\textbf{Manual Navigation (part-to-whole – p2w)}}: This user-controlled, arrow-key traversal enables fine-grained data inspection. Movement is constrained within the current segment, with distinct audio cues indicating when a boundary is reached. The directional logic of the arrow keys is relative to the active navigation axis, providing intuitive control for detailed exploration \textit{(DG2)}.
    \item [2.] \textit{\textbf{Autoplay Navigation (whole-to-part – w2p)}}: Triggered with the `P` key, this mode provides a rapid overview of the entire dataset by systematically traversing it plane-by-plane. The traversal speed is fixed at 8 points per second in Point Mode and 4 rectangles per second in Surface Mode. Furthermore, pressing the `I` key during an active autoplay session in Surface Mode activates an ``intelligent'' autoplay, which dynamically adjusts the traversal speed—slowing down in regions with significant features like peaks and troughs and speeding up over uniform areas \textit{(DG1)}.
\end{itemize}

\begin{table}[ht]
\centering \tiny
\begin{tabular}{|l|p{6cm}|}
\hline
\textbf{Key(s)} & \textbf{Function} \\
\hline
\texttt{$\uparrow / \downarrow / \leftarrow / \rightarrow$} & Navigate point-by-point within the current active segment \\
\texttt{Ctrl + Shift + $\uparrow$ / $\downarrow$} & Jump between segments along the currently active navigation axis \\
\texttt{N} & Toggle the active navigation axis for traversal (Y $\rightarrow$ Z $\rightarrow$ X) \\
\texttt{Enter} & Announce the coordinates and values of the current point or region \\
\hline
\texttt{P} & Start or stop the automated Autoplay data overview \\
\texttt{I} & During Autoplay (Surface mode only), switch to Intelligent Fast Autoplay with adaptive speed \\
\hline
\texttt{S} & Toggle sonification (audio feedback) on or off \\
\texttt{T} & Cycle through text verbosity levels for announcements (Verbose $\rightarrow$ Terse $\rightarrow$ Super-terse) \\
\texttt{F} & Cycle through preset text-to-speech (TTS) speech rates \\
\texttt{Ctrl (during TTS)} & Immediately interrupt and stop the current speech synthesis playback \\
\texttt{V} & Toggle Review Mode, which shifts focus to a text field for screen reader exploration \\
\hline
\texttt{A} & Toggle the visibility of the 3D axes and their corresponding labels \\
\texttt{H} & Open or close the Help Menu system \\
\texttt{X / Y / Z} & Announce the full label and unit for the corresponding axis \\
\hline
\texttt{Gamepad Joystick} & Navigate the dataset in 3D (mapped to keyboard arrow key logic) \\
\hline
\end{tabular}
\caption{\textbf{\textit{DIXTRAL}} Complete Key Bindings \& Input Mappings}
\label{tab:dixtral-keybindings}
\end{table}

\subsection{Accessibility Integration}
\label{subsec:accessibility}

Sonification is achieved through a multi-dimensional audio mapping where the Y-value of a data point primarily determines its pitch (scaled between 200–1200 Hz). To provide additional spatial context, the X-value is mapped to the stereo pan and oscillator type (e.g., sine, square), while the Z-value influences the tone’s duration and volume \textit{(DG1)}. Boundaries of the dataset are communicated with a distinct, pre-loaded sound.

Visual highlighting is synchronized with all navigation actions to maintain spatial consistency. The system uses a three-level highlighting architecture for clarity: the currently focused point is rendered as an enlarged magenta circle, all other points within the active navigation segment are colored white, and points outside the segment are dimmed to enhance contrast. In Surface Mode, the focused wireframe cell is highlighted as a solid, filled yellow rectangle for maximum visibility \textit{(DG1)}.

Textual output, managed by the Text and TTS Controllers, is delivered through both built-in speech synthesis and ARIA live regions for screen reader compatibility \textit{(DG3)}. The `T` key cycles through three levels of verbosity (Verbose, Terse, and Super-terse), allowing users to control the level of detail in announcements. The speech rate is adjustable with the `F` key, and any ongoing announcement can be interrupted with the `Ctrl` key.

Review Mode, activated with the `V` key, is designed for in-depth exploration with screen readers. It shifts focus from the 3D plot to a standard text area containing a cumulative log of all explored data points. This allows users to navigate the history of their exploration using standard text-based screen reader commands \textit{(DG3)}. The system preserves the user's navigation state when entering and exiting this mode, signaling the transition with distinct ascending and descending audio cues.

\textbf{\textit{DIXTRAL}} also supports full gamepad input via the MDN Gamepad API, with joystick movements normalized to match the keyboard navigation schema within the Accessibility Layer \textit{(DG2, DG4)}. Upon loading any dataset, a comprehensive set of descriptive statistics is automatically computed and displayed in a side panel. These metrics include measures of central tendency (mean, median), dispersion (standard deviation, variance, range), and distribution shape (skewness, kurtosis). Users may upload their own data in CSV or JSON formats; the system performs automatic header detection and data format normalization. All imported datasets are fully compatible with the platform's rendering, navigation, and accessibility features \textit{(DG5)}. The current visualization can be exported as a PNG image, and the underlying data can be exported in CSV or JSON format for external analysis or sharing.

%% file: chapter/5_findings-discussion.tex
\section{Findings and Discussion}
\label{sec:findings_discussion}

% objective findings corresponding to each research question defined in Introduction section.

% This section presents the findings that emerged from our co-design sessions, structured around our three guiding research questions.
The findings presented in this section are not drawn from detached analysis or post hoc interviews, but from our sustained, iterative co-design engagement with the system as it was being built and tested. We operated in a constant loop of interaction, breakdown, reflection, and redesign; through observing critical user moments such as instances of confusion, friction, discovery, or delight that surfaced when R1 and R2 engaged with \textbf{\textit{DIXTRAL}}. Throughout the four-week development cycle, we maintained detailed observation logs and reflection notes after each testing session. When a co-designer became disoriented, misinterpreted a sonification cue, or suddenly achieved insight into a data pattern, we treated those events as qualitative and methodological data points, not anecdotal feedback. We then traced those moments back to their triggering design decisions and collaboratively explored possible alternatives.

Findings were thus derived through three interdependent activities: (1) continuous system interaction by our BLV co-designers using real datasets and exploratory goals; (2) discussion and retrospective reflection between sessions, during which problems were defined and reframed; and (3) rapid iteration of the underlying architecture and interaction model, followed by verification that the change resolved the initial breakdown. This process foregrounded user reasoning as the primary analytical lens. Rather than apply standardized usability metrics, we prioritized whether the system supported our collaborators’ existing cognitive strategies and analytical goals.
As a result, the insights that followed does not claim universal generalizability; rather, they represent situated design responses to concrete barriers encountered by expert BLV researchers engaging with 3D scalar data. They illustrate how system features—such as axis switching, sonification mapping, or position preservation; evolved not from theoretical ideals, but from the real-world navigation strategies, frustrations, and breakthroughs experienced during co-use. By grounding our design logic in lived expertise and iterative problem-solving, we ensured that every finding reported here is both empirically grounded and pragmatically actionable.

Addressing RQ1, our co-design observations indicate that multimodal interaction techniques employed within \textbf{\textit{DIXTRAL}} can effectively render an accessible surface plot. When using their screen reader or the built-in TTS, R1 and R2 were able to discern information that was visually represented by the 3D plot through intuitive keybindings they recalled throughout their experience with \textbf{\textit{DIXTRAL}}. R2, who has some usable vision, struggled with using the mouse to manipulate the visual plot. Based on their suggestion, R3 introduced gamepad support, providing a more intuitive alternative for panning and zooming. This multimodal approach enabled R1 to describe the data by only listening to the plot, while R2 could synchronously explore the plot through sonification and visual highlights, which helped them orient their focus within the plot points.
In addressing RQ2, we found that combining w2p and p2w exploration is critical for BLV users. Before we implemented a w2p overview, both R1 and R2 struggled with understanding the information at a high level without resorting to point-by-point interaction, making it difficult to orient themselves within the plot. However, with an amalgamated w2p and p2w strategy, R1 and R2 could first understand the plot's general contours and then explore specific data points, leading to a better understanding. Furthermore, R2 successfully identified the surface trend for two plots with the assistance of intelligent autoplay, a task they could not accomplish with p2w navigation alone.
Regarding RQ3, our work demonstrates that an accessibility-first design can support both BLV and sighted users through shared, non-conflicting features. By adding an accessibility layer to an existing offering (WebGL), we ensured visual features were complemented, not replaced, by sonification and spoken and textual information. This proved effective, as R2 was able to leverage the information provided by their screen reader through \textbf{\textit{DIXTRAL}} in tandem with the visual highlight to gain a better understanding of the plot, showcasing a design that enhances the experience for a diverse range of users.

%% file: chapter/6_conclusion.tex
\section{Future Directions}
\label{sec:conclusion}
We have open-sourced \textit{\textbf{DIXTRAL}} to support transparency and reuse; the code is available on GitHub \footnote{\url{https://github.com/SK-143381/dixtral-assets2025-demo}}. While \textbf{\textit{DIXTRAL}} provides a foundation for multimodal 3D data exploration, several areas remain for development. The current surface mode lacks support for disjoint or sparse datasets; future work must implement more flexible rendering and interaction strategies. To date, insights have come from co-design, but formal user studies are necessary. While the co-designers brought deep lived experience and technical familiarity with both, the subject matter and system mechanism, future design cycles must include novice users and those with varied technology backgrounds to validate general usability, cognitive load, and learnability.
The system should also be extended to support additional 3D data types such as volumetric, vector field, and time-varying visualizations, to broaden applicability across domains. A VR adaptation could support embodied cognition, especially when paired with haptic feedback.

%% file: chapter/8_appendix.tex
% \appendix


% \subsection*{Researcher Visual Acuity}